\documentclass[letters]{mn2e}
\usepackage{epsfig}
\usepackage{mnras_cite}
\def\HII{\hbox{H~$\scriptstyle\rm II\ $}}

\def\nW{\,{\rm nW\,m^{-2}\,sr^{-1}}}

\def\msun{\,{\rm M_\odot}}

\def\Lya{Ly$\alpha\ $}
\def\etal{{et al.\ }}
\def\spose#1{\hbox to 0pt{#1\hss}}
\def\lta{\mathrel{\spose{\lower 3pt\hbox{$\mathchar"218$}}
     \raise 2.0pt\hbox{$\mathchar"13C$}}}
\def\gta{\mathrel{\spose{\lower 3pt\hbox{$\mathchar"218$}}
     \raise 2.0pt\hbox{$\mathchar"13E$}}}
\newcommand{\be}{\begin{equation}}
\newcommand{\ee}{\end{equation}}



\begin{document}

\title[Pop III and the NIRBE]{Population III and the near-infrared background excess}
\author[P. Madau \& J. Silk]{Piero Madau$^1$ \& Joseph Silk$^2$\\ 
1. Department of Astronomy \& Astrophysics, University of California, Santa Cruz, CA 95064,
U.S.A. (e-mail: pmadau@ucolick.org)\\
2. Astrophysics, University of Oxford, UK (e-mail: silk@astro.ox.ac.uk)}

\maketitle

\begin{abstract}
We make a critical assessment of models that attribute the recently detected
near-infrared background ``excess'' (NIRBE) to the redshifted light from
Population III objects. To supply the required 25 keV/baryon at redshift 9,
Pop III massive stars must form with an efficiency exceeding 30\% in all
``minihalos'' with virial temperatures above a few hundred kelvins: to avoid
excessive metal pollution, most of the baryons once in Pop III stars must end
up in intermediate-mass black holes (IMBHs). Gas accretion onto such IMBHs
must either be inhibited or lead to early miniquasars with steep UV/X-ray
spectra, in order not to overproduce the present-day unresolved soft X-ray
background. In the latter case (NIRBE dominated by ``X-ray quiet
miniquasars"), the total mass density of IMBHs at $z\sim 9$ must be $\gta 50$
times higher than the mass density of supermassive black holes observed today
in the nuclei of galaxies. A stellar-dominated NIRBE is less economical
energetically: $\gta 5\%$ of all baryons in the universe must be
processed into Population III stars. We survey various aspects of the Population III 
hypothesis for the NIRBE, and show that the ionizing photon budget required to
account for the NIRBE is much larger than that required to explain the high
electron scattering optical depth measured by the {\it WMAP} satellite. 
\end{abstract}

\section{Introduction}

Independent measurements based on {\it COBE}/DIRBE and {\it IRTS}/NIRS data 
provide evidence for an excess isotropic emission of extragalactic origin
in the wavelength band 1.2 to 4 $\mu$m (Dwek \& Arendt 1998; Gorjian \etal 2000; Wright
\& Reese 2000; Wright 2001; Cambresy \etal 2001; Matsumoto \etal 2004). The
inferred flux appears too bright to be accounted for by the integrated light
from faint galaxies, and shows a spectral discontinuity from the resolved
optical background at $\sim 1\mu$m (Wright \& Reese 2000; Cambresy \etal
2001; Matsumoto \etal 2004). A similar conclusion is obtained from the
analysis of near-infrared background (NIRB) fluctuations on angular scales
ranging from half a degree (DIRBE) to sub-arcmin (2MASS), namely that normal
galaxies cannot account for the observed angular power spectrum (Kashlinsky \&
Odenwald 2000; Kashlinsky \etal 2002).  It has been sugggested that the
observed NIRB spectral excess and its brightness fluctuations may be the
signature of the redshifted UV light from Population III (zero-metallicity)
star formation at $z\sim 10$ (Santos \etal 2002; Salvaterra \& 
Ferrara 2003; Magliocchetti \etal 2003; Kashlinsky \etal 2004; 
Cooray \etal 2004). Alternative possibilities involve Eddington-limited
emission from an early population of accreting miniquasars (Cooray \& Yoshida
2004), or, perhaps less exotically, flux missed by standard photometry in the
outer, lower surface brightness parts of galaxies (Bernstein \etal 2002).

In this paper we assess the merits of Pop III objects as the root cause of 
the NIRB excess (NIRBE). We show that the energetic requirements are uncomfortably 
high, and discuss the astrophysical implications of an early epoch of  massive 
Pop III star formation. 

\section{A Pop III model for the NIRBE}
\subsection{Simple energetics}

In the near IR, out of $\sim 35\,\nW$ of integrated flux from 1 to 4 $\mu$m,
about $I_J=13.5\pm 4.2\,\nW$ are contributed by the Cambresy \etal (2001)
DIRBE measurement in the J-band at 1.25 $\mu$m.\footnote{Throughout this
paper we will quote surface brightness values integrated over the bandwidth,
rather than the more common brightness per logarithmic bandwidth.}\ 
This number is quite uncertain, as it depends on zodiacal light subtraction (Kelsall 
\etal 1998; Wright 1998). Wright's zodiacal light model yields a lower value 
for the J-band NIRB,
$I_J=7.2\pm 4.0\,\nW$ (Wright 2001). The contribution from known sources
obtained by integrating the deepest galaxy counts at 1.1 $\mu$m amounts to
$2.6^{+0.8}_{-0.5}\,\nW$ (Madau \& Pozzetti 2000).
The leftover light that is often attributed to Pop III stars then ranges from a few to
about $10\,\nW$.  {\it To be conservative, we shall adopt in the following a
representative value for the NIRBE in the J-band of $2.5\,\nW$}. This
corresponds to an observed excess energy density of 
\be
u_J=10^{-15}\,{\rm ergs\, cm^{-3}}.
\ee
Assume, for simplicity, that all this energy was
emitted as \Lya radiation\footnote{This is energetically advantageous, since
most of the power of massive Pop III stars is emitted above 1 ryd, and in
Case B recombination approximately 50\% of this energy is re-emitted as
Ly$\alpha$.}\ by very massive Pop III stars with a narrow redshift
distribution centered around $\langle z\rangle=9$.  This is the redshift at
which the data requires a rapid transition from Pop III to normal Population
II stars (Salvaterra \& Ferrara 2003): a spectral jump in the NIRB 
then occurs at the redshifted wavelength $(1+\langle z\rangle)\,1216\,$\AA.  The
energetics would be even more severe if the sources responsible for the NIRBE
were broadly distributed in the range $z=9-20$ (say), or if the initial
stellar mass function (IMF) was not dominated by very massive stars.
The comoving \Lya energy density radiated at $\langle z\rangle =9$ is then
\be
U_\alpha=(1+\langle z\rangle)u_J=10\,u_J, 
\ee
corresponding to 25 keV per cosmic baryon.  Zero-metallicity stars in the range 
$80<m_*<500\,\msun$ emit $N_\gamma=70,000-90,000$ photons 
above 1 ryd per stellar baryon over a lifetime of $t_{\rm ms}=2-3\times
10^6\,$yr (Schaerer 2002). In the following, we will adopt a mid-range value of
$N_\gamma=80,000$ ionizing photons per stellar baryon. In the Case B
approximation, a fraction 0.7 of these is converted into \Lya.
The total mass density of baryons that needs to be processed by
$\langle z\rangle =9$ into Pop III stars to produce the NIRBE is
then simply estimated as \be \rho_*={m_pU_\alpha\over 0.7 N_\gamma
E_\alpha}\approx 2.7\times 10^8\, {\rm M_\odot\, Mpc^{-3}},
\label{eqrho*}
\ee where $m_p$ is the proton mass and $E_\alpha=10.2\,$eV is the energy
corresponding to the \Lya transition. Note that this mass density depends
only weakly on the IMF of massive ($>80\,\msun$) Pop III stars. Dividing
$\rho_*$ by the critical density one gets\footnote{Throughout this paper we adopt a flat
$\Lambda$CDM background cosmology with parameters ($\Omega_\Lambda, \Omega_M,
\Omega_b, n, \sigma_8, h)=(0.3, 0.7, 0.045, 1, 0.9, 0.7)$.}
\be
\Omega_*=0.002=0.045\,\Omega_b,
\label{eqOIII}
\ee
i.e. {\it at least 5\% of all baryons in the universe must be converted
into Pop III stars} (see also Santos \etal 2002). This is energetically and
astrophysically daunting. It means Pop III star formation at early times must
be comparable with the mass density in stars observed today, $(2.3\pm 0.3)
\times 10^{-3}$ (for a Kennicutt IMF; Cole \etal 2001).
    
\subsection{Star formation efficiency}

\begin{figure}
\vspace{-0.2cm}
\centerline{\epsfig{figure=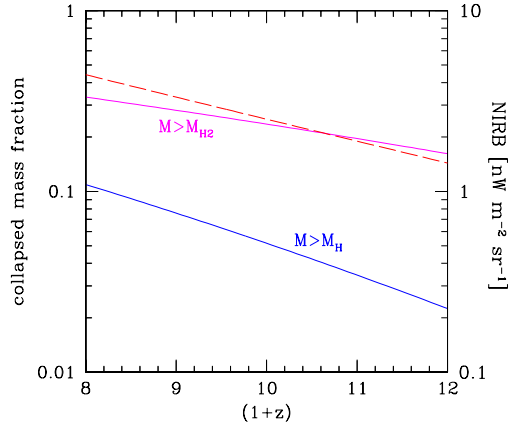,width=2.7in}}
\caption{{\it Solid lines:} Total mass fraction in all collapsed dark 
matter halos above the molecular cooling and the atomic cooling masses, $M_{\rm H_2}$ 
and $M_{\rm H}$, as a function of redshift. {\it Dashed line:} NIRB ($\nW$) due to 
redshifted \Lya emission from Population III stars at redshift $z$ (for $f_*=0.3$, 
$g=0.65$, $N_\gamma=80,000$, and $F(>M_{\rm H_2}|z)$, see text for details).}
\label{fig1}
\end{figure}

The minimum mass scale for the gravitational aggregation of cold dark matter
particles is negligibly small. In the Press-Schechter formalism, the mass
fraction in dark matter halos above mass $M$ at redshift $z$ can be written
as \be F(>M|z)={\rm erfc}\left[{\delta_c(z)\over \sqrt{2}\sigma_M(z)}\right],
\ee where $\sigma_M(z)$ is the linear theory rms density fluctuation smoothed
with a `top-hat' filter of mass $M$ at redshift $z$, and $\delta_c(z)$ is the
critical threshold on the linear overdensity for spherical collapse at that
redshift (Press \& Schechter 1974). Baryons, however, respond to pressure 
gradients and need to
contract and cool in order to fragment into stars. It is useful here to
identify two mass scales: (1) a {\it molecular cooling mass} $M_{\rm H_2}$
above which gas can cool via roto-vibrational levels of H$_2$ and contract,
$M_{\rm H_2}\approx 10^5\, [(1+z)/10]^{-3/2}\,\msun$ (virial temperature
in excess of $200\,$K; Machacek \etal 2003); and (2) {\it an atomic cooling mass} 
$M_{\rm H}$ above which
gas can cool efficiently and fragment via excitation of hydrogen Ly$\alpha$,
$M_{\rm H} \approx 10^8\, [(1+z)/10]^{-3/2}\,\msun$ (virial temperature above
$10^4\,$K). Figure \ref{fig1} shows the fraction of the total mass in the
universe that is in collapsed dark matter halos with masses greater than
$M_{\rm H_2}$ and $M_{\rm H}$ at the epochs of interest here.  Let us define
a star formation efficiency $f_*$ as the fraction of collapsed baryons that
cool and form Pop III stars, 
\be f_*={\Omega_*\over
\Omega_b\, g\, F(>M_{\rm H_2}|9)}.
\ee 
Here $g\le 1$ is a correction factor
accounting for the fact that the gas fraction (gas mass/virial mass) in
``minihalos'' condensing due to H$_2$ cooling may be lower than the universal
cosmic value $\Omega_b/\Omega_M$. In the absence of a UV photodissociating
flux and of ionizing X-ray radiation, three-dimensional simulations of early
structure formation yield gas fractions $g=0.6-0.7$ (Machacek \etal 
2003). The same simulations also show that the amount of cold, dense gas
available for star formation exceeds 20\% only for halos above $10^6\,\msun$.

From equation \ref{eqOIII} and Figure \ref{fig1}, it is obvious that if the whole
J-band excess is caused by Pop III stars, then one needs to invoke 
high star formation efficiencies, $f_*\approx 0.3$, {\it in all halos above the 
molecular cooling mass}. Pop III star formation that was limited only to halos above the
atomic cooling mass could make a significant contribution to the NIRBE only for
$f_*\sim 1$, which is implausible.
Our estimate of $f_*$ agrees with that of Santos \etal (2002) but is higher 
than the value of 10\% derived by Salvaterra \& Ferrara (2003) for a very top-heavy IMF. 
While comparably high efficiencies appear to be needed for globular clusters to survive 
gas expulsion and remain gravitationally bound (Goodwin 1997),
it is not clear how such large values may be reached in metal-free gas restricted to 
relatively inefficient H$_2$ cooling. It is also surprising that negative radiative, 
mechanical, and chemical ``feedback'' effects would not be choking off the formation 
of such a large number of Pop III stars, given the fragile nature of these 
molecules (Haiman \etal 2000; Ricotti \etal 2002; Scannapieco \etal 2002).  

\subsection{Reprocessing of ionizing radiation}

The interpretation of the J-band NIRBE as redshifted \Lya from $\langle z\rangle =9$ 
requires the emission of $U_\alpha/(0.7 E_\alpha\,n_b)\approx 3,500$
Lyman continum photons per baryon in the universe. Here $n_b=2.5\times 
10^{-7}\,$cm$^{-3}$ is the mean cosmological baryon density today. 
The assumption that most of this radiation gets absorbed and converted into \Lya within 
the host halo then implies that every baryon in the halo that was not converted into
stars must recombine on average $N_{\rm rec}=N_\gamma\,f_*/(1-f_*)\approx 
35,000$ times over the 
lifetime of a Pop III star cluster. The volume-averaged hydrogen recombination 
timescale for halo ionized gas at temperature $10^4$ K with mean overdensity 
$\delta=200\,\delta_{200}$ relative 
to the background baryon density is 
\be
t_{\rm rec}\approx 3\times 10^6\,{\rm yr}\, \left(1+z\over 10\right)^{-3}\,
\delta_{200}^{-1}\,C_h^{-1}, 
\ee
where $C_h\equiv \langle n^2_p\rangle/\langle n_p \rangle^2$ takes into account the 
degree of clumpiness of photoionized halo gas (here $\langle n_p \rangle$ is the mean 
number density of ionized hydrogen in the halo). Then from $N_{\rm rec}=(t_{\rm 
ms}/t_{\rm rec})\approx [(1+z)/10]^3\,\delta_{200}\,C_h,$
one derives a clumping factor $C_h\sim N_{\rm rec}$, i.e. the Lyman continuum photons 
must be absorbed in the densest molecular clouds around the sites of star formation
(see also Santos \etal 2002).
A small ionizing-photon escape fraction, $f_{\rm esc}$, from minihalos into the 
IGM is in contrast with the recent 
results of 1-d radiation hydrodynamical calculations of the evolution of \HII regions 
around Pop III stars  (Whalen \etal 2004). These show that 
the ionization fronts 
exit the halo on timescales much shorter than the stars' main sequence lifetimes: the 
photoevaporative flows leave dark halos with a low gas content and produce escape 
fractions $f_{\rm esc}\gta 0.95$.  
If this were the case, reprocessing of ionizing radiation into \Lya would still 
occur in the low-density IGM (Santos \etal 2002; Haardt \& Madau 1996). Note that Case B 
recombination predicts approximately 10\% of the \Lya luminosity to be reprocessed into 
redshifted H$\alpha$ line emission at 6.5 $\mu$m. 

\subsection{Metal enrichment}

Non-rotating Pop III stars in the mass window $140\lta m_*\lta 260\,\msun$
will disappear as pair-instability supernovae (PISNe; Bond \etal 1984),
leaving no compact remnants and polluting the universe with heavy
elements. Stars with $40<m_*<140\,\msun$ and $m_*>260\,\msun$ are predicted
instead to collapse to IMBHs: the mass of metals (mostly oxygen) ejected by PISNe 
is approximately equal to $m_*/2$ (Heger \& Woosley 2002). Therefore
if stars giving birth to PISNe were responsible for the NIRBE, they would
overenrich the universe to a mean metallicity, $\langle Z\rangle=\Omega_Z/\Omega_b=
0.5\Omega_*/\Omega_b\approx 0.025$,
{\it in excess of solar already by redshift 9 or so}. Even allowing for only
partial mixing of the ejecta, it is difficult to see how all these metals
could remain undetected: the unusual nucleosynthetic signature of PISNe would
appear in the metal abundances of second-generation stars,\footnote{It has been 
suggested that the relative abundances in extremely metal-poor
Galactic halo stars may show evidence for PISN enrichment below metallicities
[Fe/H]$=-3$ (Qian \& Wasserburg 2002). The inferred fraction of baryons 
in the universe processed into PISN progenitors
is, however, only $\lta 10^{-4}$ (Oh \etal 2001), approximately 500 times lower than
the value needed to explain the NIRBE (see eq. \ref{eqOIII}).}\ while
absorption studies of the \Lya forest at $z\sim 3$ would reveal cosmic
metallicities far in excess (by $\gta 2$ orders of magnitude) of the observed
values (Schaye \etal 2003).
A similar argument has recently been used to strongly  constrain  the possible 
contribution of PISNe to the reionisation of the universe (Daigne et al. 2004).

\subsection{Miniquasars}

As already pointed out by Santos \etal (2002), if distant Pop III stars make
a significant contribution to the NIRBE, then perhaps the most conservative
hypothesis is to assume that most of the baryons once in Pop III stars ended
up in IMBHs. If these seed holes were able to increase their initial mass via
gas accretion and shine as miniquasars, and a significant fraction of the
radiated energy was  emitted above and close to the hydrogen Lyman edge, then
they -- and not their metal-free stellar progenitors -- would dominate the UV
background flux (Madau \etal 2004).  Let us denote with $f_{\rm UV}$ the
fraction of the bolometric power radiated by miniquasars that is emitted as
hydrogen-ionizing radiation with mean energy $\langle E\rangle$. Then the
number of Lyman continuum photons produced per accreted baryon is
$N_\gamma=\eta m_p c^2 f_{\rm UV}/\langle E\rangle$, where $\eta$ is the
radiative efficiency. Accretion via a thin disk predicts most IMBHs at early
times to be rapidly rotating with radiative efficiencies exceeding 10-15\%
(Volonteri \etal 2005).  If the shape of the emitted spectrum from
miniquasars followed the mean spectral energy distribution of the quasar
sample in Elvis \etal (1994),\footnote{This has the same 2500 \AA--2 keV spectral
slope, $\alpha_{\rm OX}=1.4$, of the Sazonov \etal (2004) template.}\
then $f_{\rm UV}=0.3$ and $\langle
E\rangle=3\,$ryd. One then estimates $N_\gamma\approx 10^6 \eta_{0.15}$, more
than 10 times larger than the typical value for Pop III stars. Under the
assumption that all these photons were converted into \Lya radiation at
$\langle z\rangle=9$, the total mass density that must be accreted onto IMBHs
to explain the NIRBE is \be \rho_{\rm acc}={m_pU_\alpha\over 0.7 N_\gamma
E_\alpha}\approx 2\times 10^7\,\eta_{0.15}^{-1}\,{\rm M_\odot\, Mpc^{-3}}.
\label{eqIMBH}
\ee In the case of accretion at the Eddington rate, the hole mass
exponentiates over a Salpeter timescale, $t_S=(4.5\times 10^8\,{\rm
yr})~\eta/(1-\eta)$. By redshift 9, the mass density of Pop III holes could
increase by as many as 6 $e$-foldings (taking $\eta=0.15$), or about a factor
400. A miniquasar model for the NIRBE may then require an initial IMBH
progenitor mass density of only $\rho_*\approx \rho_{\rm acc}/400\approx 
50,000\,{\rm M_\odot\,Mpc^{-3}}$,
$\gta 5000$ times lower than that given in equation (\ref{eqrho*}).
Note that our results differ from those presented by Cooray 
\& Yoshida (2004). These authors fix the amount of Pop III star formation occurring
at early times and then use the observed NIRBE to constrain the fraction of stellar 
mass converted into Eddington-limited seed IMBHs to be $\lta 10\%$. To avoid 
overenrichment we take the seed hole fraction to be unity instead, and allow the 
mass density of Pop III holes to exponentiate via gas accretion, thus decreasing 
the amount of baryons that needs to be converted initially into Pop III stars.

The scenario described above is, in principle, an attractive solution to the
NIRBE problem, as it is more economical energetically and does not demand
very high star formation efficiencies down to minihalo masses. Still, plenty
of cold gas must be made available at early times for IMBHs to accrete,
driving the total mass density of IMBHs at $z\sim 9$ to more than 50 times
higher than the inferred mass density of supermassive black holes today
(Yu \& Tremaine 2002). The main issue, however, is that miniquasars
powered by IMBHs are expected to be hard X-ray emitters. Hard X-ray photons
produced in the $(0.5-2.0)(1+\langle z\rangle)\,$keV band by our population
of distant miniquasars would redshift without absorption and saturate the
unresolved soft X-ray background (SXB). For a simple power-law $F_E\propto
E^{-\alpha}$ with $\alpha>0$, the relative contribution to the
NIRBE and the SXB is 
\be 
{u_J\over u_{\rm SXB}} \sim 0.5\,\left({\alpha-1\over \alpha}\right)/\left[\left({5\,
{\rm keV}\over 13.6\,{\rm eV}}\right)^{1-\alpha}-\left({20\, 
{\rm keV}\over 13.6\,{\rm eV}}\right)^{1-\alpha}\right].
\label{eqsxb}
\ee The energy density associated with the unaccounted component of the
present-day $0.5-2\,$keV SXB does not exceed $u_{\rm SXB}\approx 2\times
10^{-18}\,{\rm ergs\, cm^{-3}}$ (Dijkstra \etal 2004),
500 times smaller than the NIRBE in the J-band. Equation (\ref{eqsxb}) is
consistent with the observations only for power-laws steeper than $\alpha\gta
2.3$ (see also Salvaterra \etal 2005). While the photon energy distribution of putative 
high-$z$ miniquasars is very uncertain, the spectra of
`ultraluminous' X-ray sources in nearby galaxies appear to require both a soft
component (well fit by a cool multicolor disk blackbody with $kT_{\rm
max}\simeq 0.15\,$keV, which may indicate IMBHs; Miller \& Colbert 2004) and
a non-thermal power-law component of comparable luminosity and slope
$\alpha\sim 1$. Even if miniquasars powered by accreting IMBHs were 
so ``X-ray quiet'' as to make a dominant contribution to the NIRBE while satisfying the 
SXB constraint, only a small fraction $f_w$ of 
the energy released could be used to drive an outflow and be ultimately deposited as
thermal energy into the IGM. Assuming rapid, homogeneous ``preheating'' at redshift 
9 (Benson \& Madau 2003), the comoving thermal energy density introduced in the IGM 
would be $U_{\rm IGM}=f_w\,U_\alpha$.
At these early epochs, inverse Compton scattering would transfer all
the energy released to the cosmic microwave background (CMB), producing a
$y$-distortion to its spectrum, $y=U_{\rm IGM}/(4U_{\rm CMB})=6\times
10^{-4}\,f_w$. The {\it COBE} satellite measured $y<1.5\times 10^{-5}$
(Fixsen et al. 1996), implying $f_w<0.025$. 

\subsection{IMBHs in galaxy halos}

We are therefore led to consider the possibility of a large population of
IMBHs ``wandering'' in galaxy halos and unable to accrete significant amounts
of material over a Hubble time. As the NIRBE would now be dominated by their
stellar progenitors, the total mass density of remnant IMBHs, $\approx 0.05\Omega_b$, 
would exceed the mass density of the
supermassive variety found today in the nuclei of most nearby galaxies,
$\Omega_{\rm SMBH}=(2.1\pm 0.3)\times 10^{-6}$ (Yu \& Tremaine 2002), {\it by
3 orders of magnitude}.  The abundance of IMBHs -- assuming they behave
similarly to collisionless dark matter particles -- is a function of the
environment, being higher (positively biased) within a massive galaxy bulge
that collapsed at early times, and lower (antibiased) within a dwarf galaxy collapsing at the 
present--epoch. This effect can be quite easily quantified within the extended Press--Schechter formalism,
and is expected to be weak for progenitor halos of mass $M_{\rm H2}$ collapsing at $z=9$ and merging 
at later times into a ``Milky Way'' halo (Madau \& Rees 2001). As the host of IMBHs 
aggregate hierarchically into more massive systems, dynamical friction against the 
visible and dark matter background will favor accretion of larger mass subunits and 
infalling satellites with low initial angular momenta: their cores will merge and undergo 
violent relaxation. Most IMBHs will not sink to the center and will be left wandering in 
galaxy halos (Madau \& Rees 2001; Volonteri \etal 2003; Islam \etal 2003), where they 
can most easily escape detection.    

\section{Discussion}

The Pop III hypothesis for the NIRBE has appeared often in studies of the
extragalactic background light, the first stars and their importance for
reionization, the duration and extent of metal-free star formation (Kashlinsky 
2004). It is only fair to point out
that the very existence of this excess is open to question, because of the
difficulty of accurately subtracting the contribution of bright
foregrounds. 
After surveying many aspects of the Pop III model for the NIRBE,
it is hard to reach any firm conclusion regarding its tenability, the unknown 
initial mass function of Pop III stars being one of the many adjustable theoretical 
parameters. Most previous
studies have emphasized the possibility that both the NIRB fluctuations and
mean levels may provide direct information on the epoch of the first
stars. While a pregalactic contribution to the NIRB at a level of a few $\nW$ cannot be 
ruled out, we offer here an assessment of the main weaknesses of the Pop III
hypothesis.
For a given amount of material converted into stars or accreted onto black holes, 
a factor $(1+z)$ is lost to cosmic expansion in the total background light observed at 
Earth when converting from observed to radiated (comoving) luminosity density. An early 
epoch of production of the NIRBE thus makes the energetic requirements even more 
extreme. A very top-heavy stellar IMF helps with the energy budget but overenriches 
the universe with heavy elements unless most baryons once in Pop III stars end 
up in IMBHs. A fine tuning of the IMF is then required to avoid the PISNe mass 
window. Radiately efficient accretion onto such a large population of IMBHs will, 
however, easily overproduce the SXB. 

The recent detection by the {\it Wilkinson Microwave Anisotropy Probe} ({\it WMAP}) 
satellite 
of a large optical depth to Thomson scattering, $\tau_e=0.17\pm 0.04$, suggests that 
ionizing sources were already very abundant at redshifts $z>10$ (Spergel \etal 2003). 
If the universe is suddenly reionized at redshift $z_r$, the Thomson scattering 
optical depth to $z_r$ is $\tau_e=\int_{t(z_r)}^{t(0)} \sigma_T\, n_e(t)\, cdt$,
where $\sigma_T$ is the Thomson cross section and $n_e(t)$ is the proper 
free-electron density. The mean number of 
recombinations per hydrogen atom in the IGM since redshift $z_r$ is 
${\cal N}_{\rm rec}=\int_{t(z_r)}^{t(0)} dt/t_{\rm rec}$.
As long as the temperature of the IGM remains constant, $T=10^4\,$K, with redshift, ${\cal 
N}_{\rm rec}$ can be written as a function of $\tau_e$ alone (Dijkstra \etal 2004) as 
\be
{\cal N}_{\rm rec}={\alpha_B(T)\, C_{\rm IGM}\over \sigma_T\, c}\,\tau_e
\approx C_{\rm IGM}\, (\tau_e/0.08),
\ee where $\alpha_B$ is the recombination coefficient. Here $C_{\rm IGM}$ is the 
clumping factor of the diffuse IGM at high redshift and is expected to be of order 
unity (we purposely exclude halo gas in our definition of $C_{\rm IGM}$, as 
the consumption of ionizing radiation by halo gas is accounted for by the escape  
fraction $f_{\rm esc}$). To 
keep intergalactic gas ionized, one then needs the production of ${\cal N}_\gamma=(1+
{\cal N}_{\rm rec})/f_{\rm esc}$ Lyman continuum photons per atom in the IGM. 
A J-band excess at a level of a few $\nW$, interpreted as redshifted \Lya from 
redshift 10, implies the emission of approximately 4,500 H-ionizing photons per atom 
in the universe. Even for ionizing-photon escape fractions of order a percent or so, 
the ionizing photon budget required to account for the NIRBE is clearly much larger 
than that required to explain the {\it WMAP} results. 

\section*{Acknowledgments}

We would like to thank A. Ferrara, F. Haardt, Z. Haiman, J. Primack, and R. Salvaterra 
for informative discussions, and M. Kuhlen for 
help with Figure 1. Support for this work was provided by NASA grants NAG5-11513 
and NNG04GK85G, and by NSF grants AST-0205738 and PHY99-07949 (PM).

{}

\end{document}